\documentclass[conference]{IEEEtran}

\usepackage[cmex10]{amsmath}
\usepackage{cite}

\usepackage{latexsym}
\usepackage{graphics}
\usepackage{amssymb}
\usepackage{amsthm}

\usepackage{cite}
\usepackage{array}
\usepackage{graphicx}

\usepackage{url}

\def\tr{\mathrm{tr}}

\def\diag{\mathrm{diag}}

\newcommand{\condSum}[3]{\overset{#3}{\underset{\underset{#2}{#1}}{\sum}}}

\newcommand{\fracSumtwo}[2]{\overset{#2}{\underset{#1}{\sum}}}
\newcommand{\vect}[1]{\mathbf{#1}}

\newcommand{\maximize}[1]{{\underset{{#1}}{\mathrm{maximize}}}}

\theoremstyle{remark}
\newtheorem{remark}{Remark}

\newtheorem{theorem}{Theorem}
\newtheorem{corollary}{Corollary}
\newtheorem{lemma}{Lemma}
\newtheorem{definition}{Definition}

\begin{document}

\title{Designing Multi-User MIMO for Energy Efficiency: \\
When is Massive MIMO the Answer?}

\IEEEoverridecommandlockouts

\author{\IEEEauthorblockN{Emil Bj{\"o}rnson\IEEEauthorrefmark{1}\IEEEauthorrefmark{2}, Luca Sanguinetti\IEEEauthorrefmark{1}\IEEEauthorrefmark{3}, Jakob Hoydis\IEEEauthorrefmark{4}, and
M{\'e}rouane Debbah\IEEEauthorrefmark{1} \thanks{E.~Bj\"ornson is funded by an International Postdoc Grant from the Swedish Research Council. L.~Sanguinetti is funded by the People Programme (Marie Curie Actions) FP7 PIEF-GA-2012-330731 Dense4Green. This research has been supported by the ERC Starting Grant 305123 MORE. Parts of this work was performed in the framework of the FP7 project ICT-317669 METIS.}}
\IEEEauthorblockA{\IEEEauthorrefmark{1}Alcatel-Lucent Chair on Flexible Radio, SUPELEC, Gif-sur-Yvette, France (\{emil.bjornson, merouane.debbah\}@supelec.fr)}
\IEEEauthorblockA{\IEEEauthorrefmark{2}ACCESS Linnaeus Centre, Signal Processing Lab, KTH Royal Institute of Technology, Stockholm, Sweden}
\IEEEauthorblockA{\IEEEauthorrefmark{3}Dip. Ingegneria dell'Informazione, University of Pisa, Pisa, Italy (luca.sanguinetti@iet.unipi.it)}
\IEEEauthorblockA{\IEEEauthorrefmark{4}Bell Laboratories, Alcatel-Lucent, Stuttgart, Germany (jakob.hoydis@alcatel-lucent.com)}
}

\maketitle

\begin{abstract}
Assume that a multi-user multiple-input multiple-output (MIMO) communication system must be designed to cover a given area with maximal energy efficiency (bit/Joule). What are the optimal values for the number of antennas, active users, and transmit power? By using a new model that describes how these three parameters affect the \emph{total} energy efficiency of the system, this work provides closed-form expressions for their optimal values and interactions. In sharp contrast to common belief, the transmit power is found to \emph{increase} (not decrease) with the number of antennas. This implies that energy efficient systems can operate at high signal-to-noise ratio (SNR) regimes in which the use of interference-suppressing precoding schemes is essential. Numerical results show that the maximal energy efficiency is achieved by a massive MIMO setup wherein hundreds of antennas are deployed to serve relatively many users using interference-suppressing regularized zero-forcing precoding. \vskip-2mm
\end{abstract}

\IEEEpeerreviewmaketitle

\section{Introduction}
The design of current wireless networks (e.g., based on the Long-Term Evolution (LTE) standard) have been mainly driven by enabling high spectral efficiency due to the spectrum shortage and rapidly increasing demand for data services \cite{Tombaz2011a}. As a result, these networks are characterized by poor energy efficiency (EE) and large disparity between peak and average rates. The EE is defined as the number of bits transferred per Joule of energy and it is affected by many factors such as (just to name a few) network architecture, spectral efficiency, radiated transmit power, and circuit power consumption \cite{Tombaz2011a,Chen2011a,EARTH_D23_short}. Motivated by environmental and economical costs, \emph{green radio} is a new research direction that aims at designing wireless networks with better coverage and higher EE \cite{Chen2011a}.

In this work, we consider the downlink of a multi-user MIMO system (broadcast channel) and aim at bringing new insights on how the number $M$ of base station (BS) antennas, the number $K$ of active user equipments (UEs), and the transmit power must be chosen in order to maximize EE. As discussed in \cite{Tombaz2011a}, a precise power consumption model is crucial to obtain reliable guidelines for EE optimization. For example, the total consumption has been traditionally modeled as a linear or affine function of the transmit power \cite{EARTH_D23_short}. However, this simple model cannot be adopted in systems where $M$ might be very large as it would lead to an unbounded EE when $M \rightarrow \infty$ \cite{Bjornson2014a}. This is because the circuit power consumed by digital signal processing and analog filters for radio-frequency (RF) and baseband processing scales with $M$ and $K$. Hence, it can be taken as a constant in small multi-user MIMO systems while the variability plays a key role when modeling so-called massive MIMO systems in which $M \gg K \gg 1$ \cite{Hoydis2013a,Ngo2013a,Bjornson2014a}.

The impact of the circuit power consumption on $M$ was recently investigated in \cite{Miao2013a,Bjornson2013e,Ha2013a,Yang2013a}. In particular, in \cite{Miao2013a} the author focuses on the power allocation problem in the uplink of multi-user MIMO systems and shows that the EE is maximized when specific UE antennas are switched off. The downlink was studied in \cite{Bjornson2013e,Ha2013a,Yang2013a}, whereof  \cite{Bjornson2013e,Ha2013a} show that the EE is a concave function of $M$ and \cite{Yang2013a} shows a similar result for $K$.
Unfortunately, these behaviors are proven only using simulations that (although useful) do not provide a complete picture of how the EE is affected by different system parameters.

In this work, we aim at closing this gap and derive closed-form expressions not only for the EE-optimal $M$, but also for $K$ and the transmit power $\rho$. These expressions provide valuable design insights about the interplay between $M$, $K$, and $\rho$, and the impact of the propagation environment as well as coefficients in the power consumption model. To ensure highly reliable results, the expressions are derived using a new power consumption model that includes high-order terms that describe how the signal processing complexity in MIMO systems scales faster than linear with $M$ and $K$. The results are derived for zero-forcing (ZF) precoding, but simulations show similar results for other common precoding schemes.

\vspace{-.2cm}
\section{Problem Formulation}

\begin{figure}
\begin{center}
\includegraphics[width=.95\columnwidth]{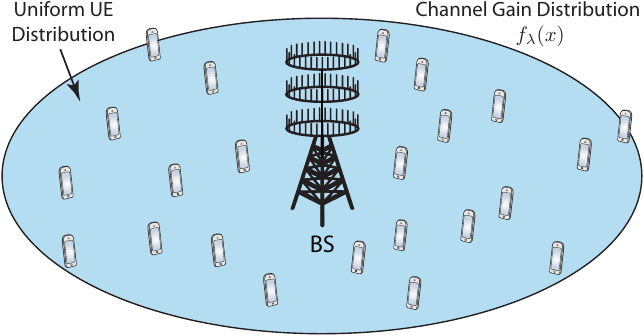}
\end{center} \vskip-5mm
\caption{Illustration of the multi-user MIMO scenario: An $M$-antenna BS transmits to $K$ single-antenna UEs. The UEs are selected randomly from a user distribution characterized by the pdf $f_{\lambda}(x)$ of the channel variances.} \label{figure_scenario} \vskip-6mm
\end{figure}

As depicted in Fig.~\ref{figure_scenario}, we consider the downlink of a multi-user MIMO system in which the BS makes use of $M$ antennas to communicate with $K$ single-antenna UEs. The flat-fading channel  $\vect{h}_{k} \in \mathbb{C}^{M}$ between the BS and the $k$th active UE is assumed to be \emph{Rayleigh block fading} as $\vect{h}_{k} \sim \mathcal{CN}(\vect{0},\lambda_{k} \vect{I}_k)$. This amounts to saying that it is constant for $T$ channel uses (c.u.) and then updated independently from the circular-symmetric complex Gaussian distribution. The $K$ active UEs change over time and are selected in a round-robin fashion from some large set of UEs that are moving around within the coverage area. For notational convenience, the active UEs are numbered as $1,2,\ldots,K$ so that the channel variances $\lambda_{1},\lambda_2,\ldots,\lambda_{K}$ are independent random variables that originate from some pdf $f_{\lambda}(x)$ describing the user distribution in the coverage area together with some location-based path loss model.

To enable acquisition of instantaneous channel state information (CSI) at the BS when $M$ is large, we consider a time-division duplex (TDD) protocol in which $K$ orthogonal uplink pilot signals are transmitted over $K$ channel uses at the beginning of each coherence block. By exploiting channel reciprocity, using Gaussian codebooks, and treating inter-user interference as noise, the average achievable information rate (in bit/channel use) of the $k$th UE is given by
\begin{equation} \label{eq:rate-expression}
R_k = \Big( 1-\frac{K}{T} \Big)  \mathbb{E} \Bigg\{ \log_2 \Bigg( 1 + \frac{ |\vect{h}_k^H \vect{v}_k|^2 }{ \condSum{\ell =1,\ell \neq k}{}{K}  |\vect{h}_k^H \vect{v}_{\ell}|^2 + \sigma^2}
  \Bigg)  \Bigg\}
\end{equation} \vskip-4mm
\noindent where the pre-log factor $(1-\frac{K}{T})$ accounts for the necessary pilot overhead and $\sigma^2$ is the noise variance. The direction $\frac{\vect{v}_k }{\| \vect{v}_k \|}$ and power $\| \vect{v}_k \|^2$ of the precoding vector $\vect{v}_k \in \mathbb{C}^{M}$ are computed on the basis of the instantaneous CSI available at the BS. The UEs have perfect CSI. The expectation in \eqref{eq:rate-expression} is taken with respect to $\{\vect{h}_{k}\}$, $\{\vect{v}_{k}\}$, and $\{\lambda_{k}\}$.

While conventional systems have large disparity between peak and average rates, we aim at designing the system so as to guarantee a uniform rate for any selected UE; that is, $R_k \!= \!R$ for some $R \!\geq\! 0$. More specifically, our goal is to \emph{find the values of $M$, $K$ and $R$ that maximize the EE of the system}.

\subsection{General Energy Efficiency Metric}

The EE of a communication system is measured in bit/Joule \cite{Chen2011a} and equals the ratio between the average achievable sum information rate (in bit/channel use) and the total average power consumption (in Joule/channel use). The power consumption of conventional macro BSs is roughly proportional to the radiated transmit power \cite{EARTH_D23_short}. However, this assumption does not hold in general. Indeed, making such an assumption in massive MIMO systems can be very misleading since an infinite EE can be achieved as $M \rightarrow \infty$ \cite{Bjornson2014a}. This calls for a more detailed and realistic model.

Apart from the power consumed by the RF power amplifier, there is a circuit power consumption from digital signal processing and analog filters used for RF and baseband processing. Inspired by the power consumption models in \cite{EARTH_D23_short,Tombaz2011a,Kim2010a,Yang2013a}, we propose a new improved model that clearly specifies how the power scales with $M$ and $K$. The total power consumption (in Joule/channel use) is\footnote{This is the power consumed by the system, but there are also losses in the power supply and due to cooling \cite{EARTH_D23_short}. These losses are typically proportional to \eqref{eq:power-consumption} and can thus be neglected in the analysis without loss of generality.}
\begin{equation} \label{eq:power-consumption}
P^{\mathrm{total}} =  \fracSumtwo{k=1}{K} \frac{\mathbb{E} \{ \| \vect{v}_k \|^2 \}}{\eta}  + \sum_{i=0}^{3} C_{i,0} K^{i}  + \sum_{i=0}^{2} C_{i,1} K^{i} M
\end{equation}
where $0 < \eta \leq 1$ is the efficiency of the power amplifier and the expectation is taken with respect to channel realizations and user locations. The term $C_{0,0} \geq 0$ is the static hardware power consumption that does not scale with $M$ or $K$.
The remaining power consumption terms are of the structure $C_{i,j} K^{i} M^{j}$ for some coefficient $C_{i,j} \geq 0$ and different integer values on $i$ and $j$. The range of high-order terms is motivated in Section \ref{subsec:example-parameters}.
We can now define our EE metric.
\begin{definition}
The average energy efficiency (EE) is
\begin{equation} \label{eq:EE-metric}
\mathrm{EE} =  \frac{ \sum_{k=1}^{K} R_k }{ P^{\mathrm{total}} }
\end{equation}
where $R_k$ and $P^{\mathrm{total}}$ are given in \eqref{eq:rate-expression} and \eqref{eq:power-consumption}, respectively.
\end{definition}

\subsection{Example: Parameters in the Power Consumption Model}
\label{subsec:example-parameters}

To motivate \eqref{eq:power-consumption}, we give a brief summary of different aspects that contribute to the total downlink power consumption.

\subsubsection{Transceiver Chains}

The typical MIMO transceivers in \cite{Cui2004a} have a power consumption of $M P_{\mathrm{tx}} + K P_{\mathrm{rx}} + P_{\mathrm{syn}} $ Joule/channel use. $P_{\mathrm{tx}}$ is the power of the BS components attached to each antenna: converters, mixers, and filters. A single oscillator with power $P_{\mathrm{syn}}$
is used for all BS antennas. Since we consider single-antenna UEs, $P_{\mathrm{rx}}$ is the power of all receiver components: amplifiers, mixer, oscillator, and filters.

\subsubsection{Coding and Decoding}

The BS applies channel coding and modulation to $K$ sequences of information symbols and each UE applies some suboptimal fixed-complexity algorithm to decode its own sequence. Therefore, the power consumption is $K ( P_{\mathrm{cod}} +P_{\mathrm{dec}})$ Joule/channel use, where $P_{\mathrm{cod}}$ and $P_{\mathrm{dec}}$ are the coding and decoding powers, respectively.

\subsubsection{Channel Estimation and Precoding}

Let the computational efficiency be  $L$ operations per Joule at the BS (also known as flops/Watt).
The uplink CSI estimation consists of receiving $M$ signals per UE and scaling each one by a factor that depends on the estimator. Since estimation takes place once per coherence period, it uses $\frac{MK}{LT}$ Joule/channel use.

The precoding is precomputed once per coherence period. Computing maximum ratio transmission (MRT) \cite{Bjornson2014a} costs $\frac{2MK}{LT}$ Joule/channel use (due to normalization), while ZF and regularized ZF cost $\frac{3 K^2 M + 2 K M}{LT}+ \frac{2K^3}{3LT}$ Joule/channel use (due to LU-based matrix inversion) \cite{Boyd2008a}. During data transmission, the precoding matrix is multiplied with the vector of information symbols, which costs $(1 - \frac{K}{T})\frac{MK}{L}$ Joule/channel use.

\subsubsection{Architectural Costs}

The system architecture incurs a fixed power consumption $P_0$ that does not scale with neither $M$ nor $K$. This term can, for example, include the fixed power consumption of control signaling, backhaul infrastructure, and the load-independent consumption of baseband processors.

\subsubsection{Summary}

The hardware characterization above gives a power consumption of the form in \eqref{eq:power-consumption}. The coefficients under ZF precoding are $C_{0,0} =  P_0 + P_{\mathrm{syn}} $, $C_{1,0} = P_{\mathrm{cod}} +P_{\mathrm{dec}} + P_{\mathrm{rx}}$, $C_{2,0} = 0$, $C_{3,0} = \frac{2}{3LT}$, $C_{0,1} = P_{\mathrm{tx}}$, $C_{1,1} = \frac{3+T}{LT}$, $C_{2,1} = \frac{2}{LT}$.

\section{Optimal EE with Zero-Forcing Precoding}

Next, we optimize the EE metric in \eqref{eq:EE-metric} for the realistic case of $M \geq K$ and under the simplifying assumptions of ZF precoding and that the pilot signaling provides the BS with perfect CSI. The following result is found.

\begin{lemma} \label{lemma:ZF-precoding}
Consider a channel realization $\vect{H} = [\vect{h}_1 \, \vect{h}_2 \ldots \vect{h}_K]$ and assume that ZF precoding is used to give each active UE an information rate of $ \big( 1-\frac{K}{T} \big) \log_2 \big(1+ \rho (M-K) \big)$ for some (normalized) transmit power $\rho>0$.\footnote{The effective channel is $\vect{h}_k^H \vect{v}_k = \sqrt{\rho \sigma^2  (M-K)}$ irrespective of the channel realizations, thus the UE needs no instantaneous CSI for decoding.} The total transmit power is $\sum_{k=1}^{K} \|\vect{v}_k \|^2 = \rho (M-K) \tr \big( (\vect{H}^H \vect{H})^{-1} \big)$ and the average total transmit power is
\begin{equation} \label{eq:average_power}
\sum_{k=1}^{K} \mathbb{E}\{ \|\vect{v}_k \|^2 \} = \rho K A_{\lambda}
\end{equation}
where $A_{\lambda} = \mathbb{E}\{ \frac{\sigma^2}{\lambda} \} = \int_{0}^{\infty} \frac{\sigma^2}{x}  f_{\lambda}(x) dx$ is given by the propagation environment.
\end{lemma}
\begin{IEEEproof}
The given rate is achieved by the ZF precoding  $\vect{V} = [ \vect{v}_1 \, \vect{v}_2  \ldots \vect{v}_K] = \sqrt{\rho \sigma^2 (M-K)} \vect{H} (\vect{H}^H \vect{H} )^{-1}$ and use power $\tr( \vect{V}^H \vect{V} ) = \rho \sigma^2 (M-K) \tr \big( (\vect{H}^H \vect{H})^{-1} \big)$. To compute \eqref{eq:average_power}, note that $\vect{H}^H \vect{H} \in \mathbb{C}^{K \times K}$ has a complex Wishart distribution with $M$ degrees of freedom and the parameter matrix $\Lambda = \diag(\lambda_1,\lambda_2,\ldots,\lambda_K)$. By using \cite[Eq.~(50)]{Maiwald2000a}, the inverse first-order moment is $\mathbb{E}\{ \tr \big( (\vect{H}^H \vect{H})^{-1} \big)\} = \mathbb{E}\{ \frac{\tr ( \Lambda^{-1} )}{M-K} \}  = \\ \sum_{k=1}^{K} \frac{ \mathbb{E}\{ \lambda_k^{-1}\} }{M-K}$, where the remaining expectation with respect to $\lambda_k$ is the same for all $k$ and computed using $f_{\lambda}(x)$.
\end{IEEEproof}

For ZF precoding, the EE metric in \eqref{eq:EE-metric} reduces to
\begin{equation} \label{eq:EE-metric-ZF}
\mathrm{EE} = \frac{ K \big( 1-\frac{K}{T} \big) \log_2 \big(1+ \rho (M-K) \big) }{ \frac{\rho K A_{\lambda}}{\eta}  + \sum_{i=0}^{3} C_{i,0} K^{i}  + \sum_{i=0}^{2} C_{i,1} K^{i} M }.
\end{equation}
This tractable expression is used herein to compute the values of $M$, $K$, and $\rho$ that maximize the EE. Although ZF precoding is highly suboptimal at low SNRs, we will show by simulation that this is not the optimal operating regime and that the guidelines derived in this section have general implications.

\begin{remark} \label{remark:propagation-environment}
The user distribution and propagation environment is characterized by $A_{\lambda}$ in Lemma \ref{lemma:ZF-precoding}. As an example, suppose the UEs are uniformly distributed in a circular cell with radius $d_{\max}$ and minimum distance $d_{\min}$. Let $\lambda = \frac{D}{d^{\kappa}}$ where $D>0$ is the fixed channel attenuation, $d$ is the distance from the BS, and $\kappa > 0$ is the path loss exponent. It is straightforward to show that $A_{\lambda} = \mathbb{E}\{ \frac{\sigma^2}{\lambda} \} = \frac{\sigma^2 }{D ( 1 + \frac{\kappa}{2})} \frac{d_{\max}^{\kappa+2} - d_{\min}^{\kappa+2}}{d_{\max}^2 - d_{\min}^2}$.
\end{remark}

\subsection{Preliminaries}

The Lambert W function appears repeatedly in this work.

\begin{definition}
The Lambert W function $W(x)$ is defined by the equation $x = W(x) e^{W(x)}$ for any $x \in \mathbb{C}$.
\end{definition}

The following lemma is of main importance.

\begin{lemma} \label{lemma:optimization-EE}
Consider the optimization problem
\begin{equation} \label{eq:lemma-opt-problem}
\maximize{ z > -\frac{a}{b}} \quad \frac{f \log_2(a+bz) }{c+d z}
\end{equation}
with constants $a \in \mathbb{R}$, $c \geq 0$, and $b,d,f>0$. The objective function is strictly quasi-concave and \eqref{eq:lemma-opt-problem} has the unique solution
\begin{equation} \label{eq:lemma-opt-problem-solution}
z^{\mathrm{opt}} = \frac{e^{W \left(\frac{bc-ad}{de} \right)+1}-a}{b}
\end{equation}
where $e$ is the natural number.
 The objective function is increasing for $z < z^{\mathrm{opt}}$ and decreasing for $z > z^{\mathrm{opt}}$.
\end{lemma}
\begin{IEEEproof}
The proof is given in the appendix.
\end{IEEEproof}

The following lemma (based on inequalities in \cite{Hoorfar2008a}) brings some insight on how the optimal solution $z^{\mathrm{opt}}$ in \eqref{eq:lemma-opt-problem} behaves.

\begin{lemma} \label{lemma:lambert-bounds}
The Lambert W function $W(x)$ satisfies $W(0) = 0$, is increasing for $x \geq 0$, and fulfills the inequalities
\begin{equation}
\frac{x \, e}{\log_e(x)} \leq  e^{W(x)+1} \leq \frac{x}{\log_e(x)} (1+e) \quad \textrm{for all} \,\, x \geq e.
\end{equation}
\end{lemma}

This implies that $e^{W(x)+1}$ is approximately equal to $e$ for small $x$ and increases almost linearly with $x$ for large $x$.

\subsection{Optimal System Parameters}

Next, we find the value of either $M$, $K$, or $\rho$ that maximizes the EE metric, when the other two parameters are fixed.

\subsubsection{Optimal Number of BS Antennas}

The optimal value of $M$ is provided by the following theorem.

\begin{theorem} \label{theorem-optimal-M}
The EE optimization problem
\begin{equation} \label{eq:theorem-opt-problem-M}
\maximize{ M \geq K} \quad \frac{ K \big( 1-\frac{K}{T} \big) \log_2 \big(1+ \rho (M-K) \big) }{ \frac{\rho K A_{\lambda}}{\eta}  + \sum_{i=0}^{3} C_{i,0} K^{i}  + \sum_{i=0}^{2} C_{i,1} K^{i} M }
\end{equation}
is solved by
\begin{equation}  \label{eq:antenna-scaling}
M^{\mathrm{opt}} \!=\! \frac{e^{W \! \left(\frac{  \frac{ \rho^2 K A_{\lambda}}{\eta}  + \rho \sum_{i=0}^{3} C_{i,0} K^{i} }{e \sum_{i=0}^{2} C_{i,1} K^{i}} + \frac{ K \rho - 1 }{e } \right)+1}\!\!+K \rho - 1}{\rho}.
\end{equation}
\end{theorem}
\begin{IEEEproof}
Follows from Lemma \ref{lemma:optimization-EE} for $a=1-K \rho$, $b = \rho$, $c= \frac{ \rho K A_{\lambda}}{\eta}  + \sum_{i=0}^{3} C_{i,0} K^{i}$, $d = \sum_{i=0}^{2} C_{i,1} K^{i}$, and $f= K \big( 1-\frac{K}{T} \big)$. $M^{\mathrm{opt}}$ is in the feasible set $K \leq M < \infty$ since the objective is  quasiconcave and equals zero at $M=K$ and $M \rightarrow \infty$.
\end{IEEEproof}

This theorem provides an explicit guideline on how many antennas should be used at the BS to maximize the EE. Using Lemma \ref{lemma:lambert-bounds} we have that: 1) $M$ increases sublinearly with the (normalized) transmit power $\rho$ but almost linearly when $\rho$ is large; 2) $M$ increases with the circuit coefficients $C_{i,0}$ that are independent of $M$ and decreases when increasing the circuit coefficients $C_{i,1}$ that are multiplied with $M$ in the EE metric; 3) $M$ increases almost linearly with $A_{\lambda}$. Recall from Remark~\ref{remark:propagation-environment} that $A_{\lambda}$ is proportional to $d_{\max}^{\kappa}$ in circular cells, where $d_{\max}$ is the radius and $\kappa$ is the path loss exponent.

Note that Theorem \ref{theorem-optimal-M} typically gives a non-integer value on $M^{\mathrm{opt}}$, but the quasiconcavity of the problem \eqref{eq:theorem-opt-problem-M} implies that the optimal $M$ is attained at one of the two closest integers.

\subsubsection{Optimal Transmit Power}

The transmit power is $\rho K A_{\lambda}$ and the optimal $\rho$ is given by the next theorem.

\begin{theorem} \label{theorem-optimal-rho}
The EE optimization problem
\begin{equation} \label{eq:lemma-opt-problem-rho}
\maximize{ \rho \geq 0}  \quad \frac{ K \big( 1-\frac{K}{T} \big) \log_2 \big(1+ \rho (M-K) \big) }{ \frac{\rho K A_{\lambda}}{\eta}  + \sum_{i=0}^{3} C_{i,0} K^{i}  + \sum_{i=0}^{2} C_{i,1} K^{i} M }
\end{equation}
is solved by
\begin{equation} \label{eq:power-scaling}
\rho^{\mathrm{opt}} \!=\! \frac{e^{W \! \left(\frac{(M-K) \eta (\sum_{i=0}^{3} C_{i,0} K^{i}  + \sum_{i=0}^{2} C_{i,1} K^{i} M)}{ K A_{\lambda} e} - \frac{1}{e} \right)+1}\!\!-1}{M-K}. \!\!
\end{equation}
\end{theorem}
\begin{IEEEproof}
Follows from Lemma \ref{lemma:optimization-EE} for $a=1$, $b = M-K$, $c=\sum_{i=0}^{3} C_{i,0} K^{i}  + \sum_{i=0}^{2} C_{i,1} K^{i} M $, $d= \frac{K A_{\lambda}}{\eta}$, and $f= K \big( 1-\frac{K}{T} \big)$. The value $\rho^{\mathrm{opt}}$ is always positive since the objective is quasiconcave and equals zero at $\rho=0$ and when $\rho \rightarrow \infty$.
\end{IEEEproof}

This theorem provides the transmit power $\rho^{\mathrm{opt}}  K A_{\lambda}$ that maximizes the EE. Recall from Lemma \ref{lemma:lambert-bounds} that $e^{W (x) +1}$ is monotonically increasing with a sublinear slope that becomes almost linear when $x$ is large. Consequently, \eqref{eq:power-scaling} shows that the optimal transmit power (and the SINR $\rho(M-K)$)
increases with the circuit powers (i.e., the coefficients $C_{i,j}$). This might seem counterintuitive but makes much sense: if the fixed circuit power is large we can afford more transmit power before that it has a non-negligible impact on the total power consumption.

It has recently been shown in \cite{Hoydis2013a,Ngo2013a,Bjornson2014a} that massive MIMO systems permit a power reduction proportional to $1/M$ (or $1/\sqrt{M}$ with imperfect CSI) while maintaining non-zero UE rates as $M \rightarrow \infty$. Although this is a remarkable result, Theorem \ref{theorem-optimal-rho} shows that this is \emph{not} the most energy efficient strategy in practice. In fact, the EE metric is generally maximized by more-or-less the opposite strategy (i.e., increase $\rho$ with $M$).

\begin{corollary} \label{cor:increase-power-with-M}
For large $M$, the optimal value in \eqref{eq:power-scaling} satisfies
\begin{equation}
\begin{split}
\rho^{\mathrm{opt}}
&\geq \frac{(\tilde{C}_0  + \tilde{C}_1 M) - \frac{\log_e ( (M-K) (\tilde{C}_0  + \tilde{C}_1 M) -1 )}{M-K}  }{ \log_e ( (M-K) (\tilde{C}_0  + \tilde{C}_1 M) -1 ) -1} \\ &= \begin{cases}
\mathcal{O} \left( \frac{M}{\log_e (M)} \right), & \tilde{C}_1>0, \\
\mathcal{O} \left( \frac{1}{\log_e (M)} \right), & \tilde{C}_1=0, \end{cases}
\end{split}
\end{equation}
where $\tilde{C}_0 = \frac{ \eta\sum_{i=0}^{3} C_{i,0} K^{i} }{K A_{\lambda}  }$ and $\tilde{C}_1 = \frac{\eta \sum_{i=0}^{2} C_{i,1} K^{i}}{K A_{\lambda} }$. The growth rates are stated using conventional big-$\mathcal{O}$ notation.
\end{corollary}
\begin{IEEEproof}
Follows by applying the lower bound in Lemma \ref{lemma:lambert-bounds} (which holds when $M$ is large) and some simple algebra.
\end{IEEEproof}

This corollary reveals that the transmit power should increase almost linearly with $M$ to maximize the EE metric. The explanation is the same as above: if the circuit power grows with $M$ we can afford using more transmit power before that it becomes the limiting factor for the EE. In the special case when the circuit power is independent of the number of antennas (i.e., $C_{i,1} =0$ for all $i$), the power should instead be reduced proportional to $\log_e (M)$. This power reduction is however much slower than the linear reduction reported in \cite{Hoydis2013a,Ngo2013a}---such scalings are only obtained in the unrealistic case when there is no circuit power consumption whatsoever.

\subsubsection{Optimal Number of UEs}

The optimal $K$ is given by the next theorem. For analytical tractability, we let the total transmit power be fixed such that $\rho^{\textrm{tot}} = K \rho $ and the number of transmit antennas available per UE be fixed as $\beta = \frac{M}{K}$.

\begin{theorem} \label{theorem-optimal-K}
The EE optimization problem
\begin{equation} \label{eq:lemma-opt-problem-K}
\maximize{ K \geq 0}  \quad \frac{ K \big( 1-\frac{K}{T} \big) \log_2 \big(1+ \rho^{\textrm{tot}} (\beta-1) \big) }{ \frac{\rho^{\textrm{tot}} A_{\lambda}}{\eta}  + \sum_{i=0}^{3} C_{i,0} K^{i}  + \sum_{i=0}^{2} C_{i,1} \beta K^{i+1} }
\end{equation}
is quasiconcave and solved by a root to the quartic polynomial
\begin{equation} \label{eq:polynomial-K}
b c_3 K^4 - 2 c_3 a K^3 - (a c_2 + b c_1) K^2 - 2 b c_0 K + c_0 a
\end{equation}
where $a = \log_2 \big(1+ \rho^{\textrm{tot}} (\beta-1) \big)$, $b = \frac{a}{T}$, $c_0 = C_{0,0} + \frac{\rho^{\textrm{tot}} A_{\lambda}}{\eta}$, $c_1 = C_{1,0} + \beta C_{0,1}$, $c_2 = C_{2,0} + \beta C_{1,1}$, and $c_3 = C_{3,0} + \beta C_{2,1}$.

In the special case of $c_3 = 0$, the optimal solution is
\begin{equation} \label{eq-optimal-K-special-case}
K^{\mathrm{opt}} = \sqrt{ \left( \frac{b c_0}{a c_2 + b c_1 } \right)^2 + \frac{c_0 a}{a c_2 + b c_1}    } - \frac{b c_0}{a c_2 + b c_1 }.
\end{equation}
\end{theorem}
\begin{IEEEproof}
The objective function is obtained from \eqref{eq:EE-metric-ZF} by substituting $\rho = \frac{\rho^{\textrm{tot}}}{K} $ and $M = \beta K$. It has the structure $g(K) = \frac{aK-bK^2}{\sum_{i=0}^{3} c_i K^i}$ and is a strictly quasiconcave function since the level sets $S_{\alpha} =  \{ K \, : \, g(K) \geq \alpha \} = \{ K \, : \, \alpha \sum_{i=0}^{3} c_i K^i + b K^2 - a K \leq 0 \}$ are strictly convex for any $\alpha \in \mathbb{R}$ \cite[Section 3.4]{Boyd2004a}.
The global optimum satisfies the stationarity condition $g'(K) = 0$, which is equivalent to finding roots of \eqref{eq:polynomial-K}.
\end{IEEEproof}

There are generic closed-form expressions for the 4 roots of any quartic polynomial (such as \eqref{eq:polynomial-K}), but these are lengthy and not given here. We refer to \cite{Shmakov2011a} for a survey on root computations. The EE-maximizing root is found by testing and is generally not an integer, but the quasiconcavity of \eqref{eq:lemma-opt-problem-K} implies that the optimal $K$ is one of the two closest integers.

To gain insight, we focus on the special case of $C_{3,0} = C_{2,1} = 0$ (i.e., ignoring the power terms in \eqref{eq:power-consumption} with the highest orders) and recall that the ratio $\beta = \frac{M}{K}$ is fixed.
The optimal number of UEs in \eqref{eq-optimal-K-special-case} is a decreasing function of $C_{1,0} , C_{2,0}, C_{0,1}, C_{1,1}$, which are the coefficients of the circuit power terms that scale with $M$ and/or $K$. However, we can afford more UEs (and BS antennas) when the power consumption is dominated by terms that are independent of $M$ and $K$; that is, $K^{\mathrm{opt}}$ increases with the static hardware power $C_{0,0}$ and the propagation environment parameter $A_{\lambda}$ (which scales with the coverage area; see Remark \ref{remark:propagation-environment}).

\subsection{Joint and Sequential Optimization of $M,K,\rho$}
\label{subsec:joint-sequential}

The simple expressions in Theorems \ref{theorem-optimal-M}--\ref{theorem-optimal-K} optimize each of $M$, $\rho$, and $K$ separately when the two other parameters are fixed. Ideally, one would like to find the joint global EE-optimum ($M^*,K^*,\rho^*$). Since $M$ and $K$ are integers, finding the optimum is guaranteed by making an exhaustive search over all reasonable combinations of $M,K$ and computing the optimal power allocation in Theorem \ref{theorem-optimal-rho} for each combination.

The exhaustive algorithm is feasible for offline cell planning, but a low-complexity approach is of interest for adaptation to changes in propagation environment (i.e., user distribution and path loss model specified by $f_{\lambda}(x)$). Given an initial set ($M,K,\rho$), we can utilize standard alternating optimization:
\begin{enumerate}
\item Update the number of UEs $K$ according to Theorem \ref{theorem-optimal-K};
\item Replace $M$ by the optimal value from Theorem \ref{theorem-optimal-M};
\item Optimize the transmit power by using Theorem \ref{theorem-optimal-rho};
\item Repeat 1)--3) until convergence.
\end{enumerate}

Since the EE metric has a finite upper bound (when some $C_{i,j}$ is strictly positive) and the EE is nondecreasing in each step, the alternating algorithm is guaranteed to converge but not necessarily to a global optimum. Convergence has occurred when the integers $M$ and $K$ are not changed in an iteration.

\section{Numerical Illustrations}
\label{sec:simulations}

This section illustrates some system design guidelines obtained from Theorems \ref{theorem-optimal-M}--\ref{theorem-optimal-K}.
To compute the power consumption in a realistic way, we use the hardware characterization described in Section \ref{subsec:example-parameters}. The corresponding parameter values are inspired by \cite{EARTH_D23_short,Kumar2011a} and summarized in Table \ref{table_parameters_hardware}. We assume a uniform user distribution in a circular cell of radius $250$ m and use a typical 3GPP distance-dependent path loss model. The propagation parameter $A_{\lambda}$ is computed as in Remark \ref{remark:propagation-environment}.

\begin{table}[!t]
\renewcommand{\arraystretch}{1.3}
\caption{Simulation Parameters}
\label{table_parameters_hardware} \vskip-2mm
\centering
\begin{tabular}{|c|c||c|c|}
\hline
\bfseries Parameter & \bfseries Value & \bfseries $\!\!$ Parameter $\!\!$ & \bfseries Value\\
\hline

Cell size: $d_{\min},d_{\max}$ &  $35$ m, $250$ m &  $\eta$ & $0.3$ \\

Pathloss at distance $d$: $\lambda$ & $\frac{10^{-3.53}}{d^{3.76}}$ & $P_0$ & $2 \, \mathrm{W} \cdot S$ \\

Coherence bandwidth: $B$ & $180$ kHz   & $P_{\mathrm{syn}} $ & $ 2 \, \mathrm{W} \cdot S$ \\

Coherence time: $T$ & $32 \, \mathrm{ms}  \cdot B \cdot \mathrm{c.u.}$ & $P_{\mathrm{cod}}  $ & $ 4 \, \mathrm{W} \cdot S$ \\

Symbol time: $S$ & $\frac{1}{9 \cdot 10^6} \,\, \mathrm{s/c.u.}$ & $P_{\mathrm{dec}}  $ & $ 0.5 \, \mathrm{W} \cdot S$ \\

Operations/Joule: $L$ & $10^9$  & $P_{\mathrm{tx}} $ & $ 1 \, \mathrm{W} \cdot S$ \\

Noise variance: $\sigma^2 $ & $ 10^{-20} \, J/\mathrm{c.u.}$ & $P_{\mathrm{rx}}  $ & $ 0.3 \, \mathrm{W} \cdot S$ \\

\hline
\end{tabular} \vskip-3mm
\end{table}

\begin{figure}
\begin{center}
\includegraphics[width=\columnwidth]{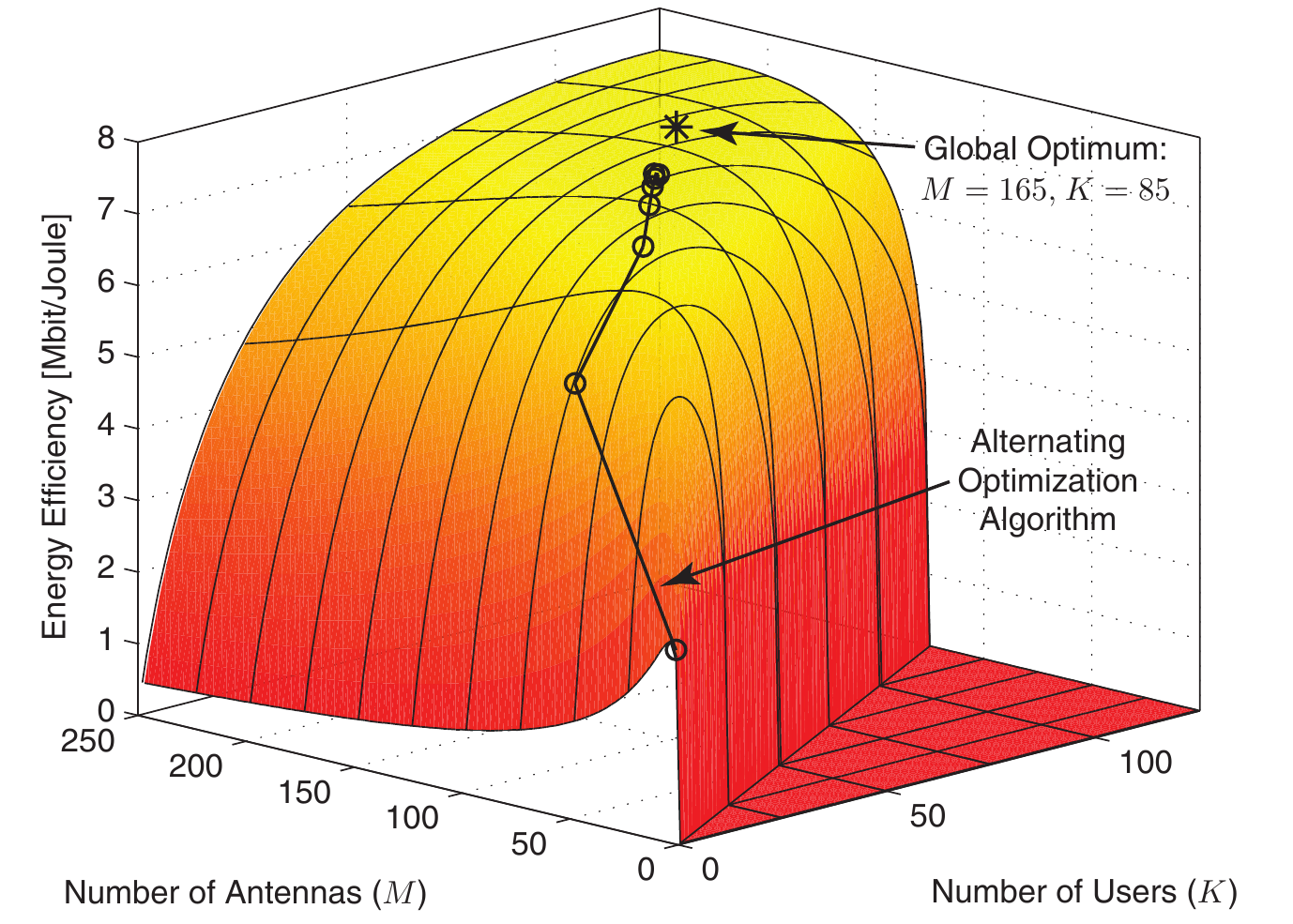}
\end{center} \vskip-5mm
\caption{Energy efficiency (in Mbit/Joule) with ZF precoding for different combinations of $M$ and $K$ (with the optimal $\rho$ from Theorem \ref{theorem-optimal-rho}). The global optimum is marked with a star, while the convergence of the alternating optimization algorithm from Section \ref{subsec:joint-sequential} is indicated with circles.} \label{figure-3dplot-zf} \vskip-4mm
\end{figure}

Fig.~\ref{figure-3dplot-zf} shows the set of achievable EE values under ZF precoding and different values of $M$ and $K$ (note that $M \geq K$ due to ZF). Each point uses the EE-maximizing value of $\rho$ from Theorem \ref{theorem-optimal-rho}. The figure shows that there is a global optimum at $M = 165$ and $K = 85$ (with $\rho=4.6097$), which we interpret as being a massive MIMO setup. The surface in Fig.~\ref{figure-3dplot-zf} is concave and quite smooth; thus, there is a variety of system parameters that provide close-to-optimal EE and the results appear to be robust to small changes in the circuit power coefficients. The alternating algorithm from Section \ref{subsec:joint-sequential} was applied with a starting point in $M=3$, $K=1$, and $\rho=1$. The algorithm converged after 7 iterations to a suboptimal solution in the vicinity of the global optimum.

\begin{figure}
\begin{center}
\includegraphics[width=\columnwidth]{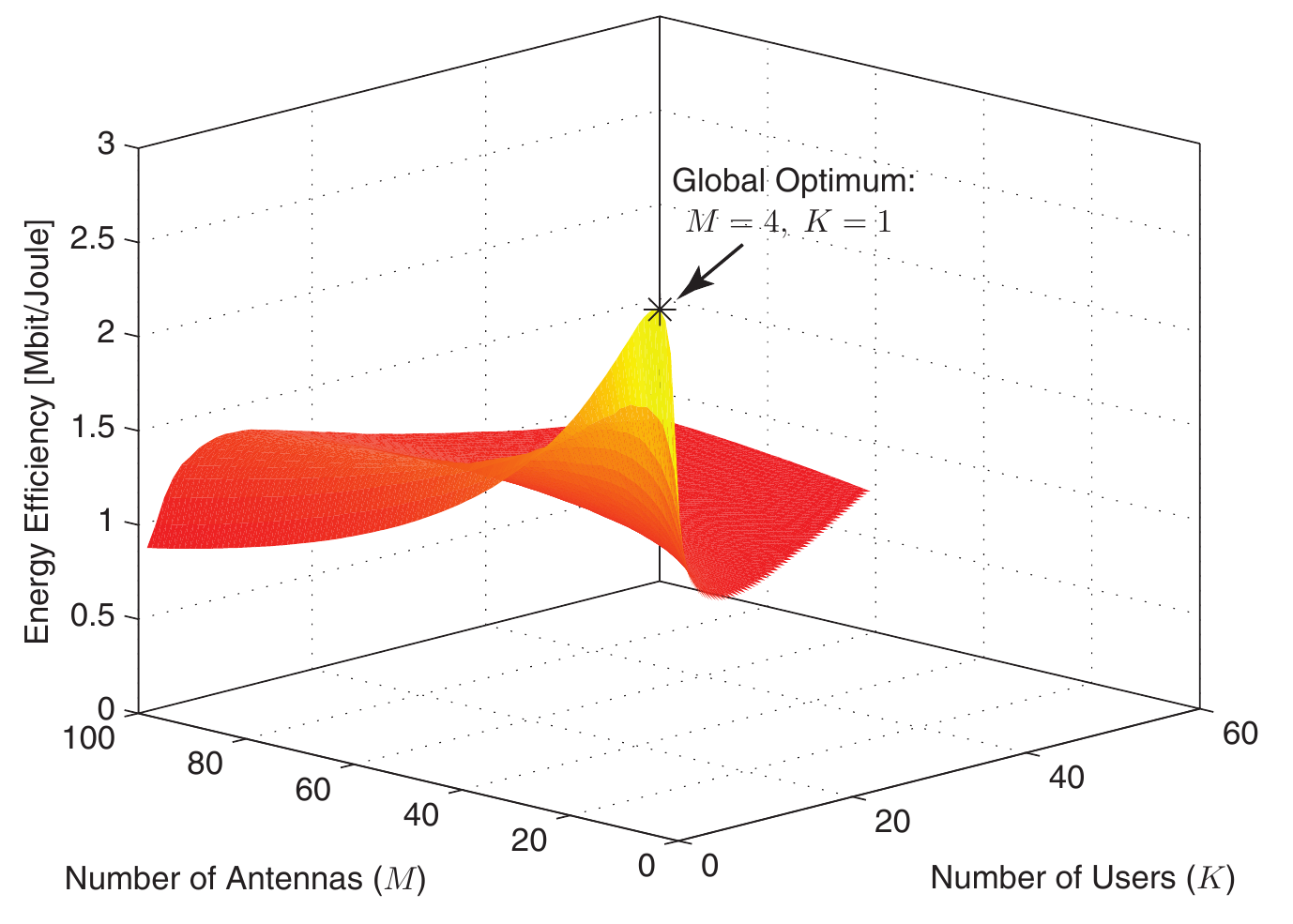}
\end{center} \vskip-5mm
\caption{Energy efficiency (in Mbit/Joule) with MRT for different combinations of $M$ and $K$ (with the optimal $\rho$ computed by MC simulations). The global optimum is marked with a star.} \label{figure-3dplot-mrt} \vskip-4mm
\end{figure}

As a comparison, Fig.~\ref{figure-3dplot-mrt} shows the corresponding set of achievable EE values under MRT. Fig.~\ref{figure-3dplot-mrt} was generated by Monte Carlo (MC) simulations, while Fig.~\ref{figure-3dplot-zf} was computed using our analytic results. Interestingly, Fig.~\ref{figure-3dplot-mrt} shows a completely different behavior: the highest EE is achieved using a small number of BS antennas and only one active UE, which is due to strong inter-user interference. Such interference reduces with the number of BS antennas since the channels decorrelate as $M \rightarrow \infty$ \cite{Hoydis2013a}, but this effect is not dominating over the increased cost in computational/circuit power of increasing $M$.

\begin{figure}
\begin{center}
\includegraphics[width=\columnwidth]{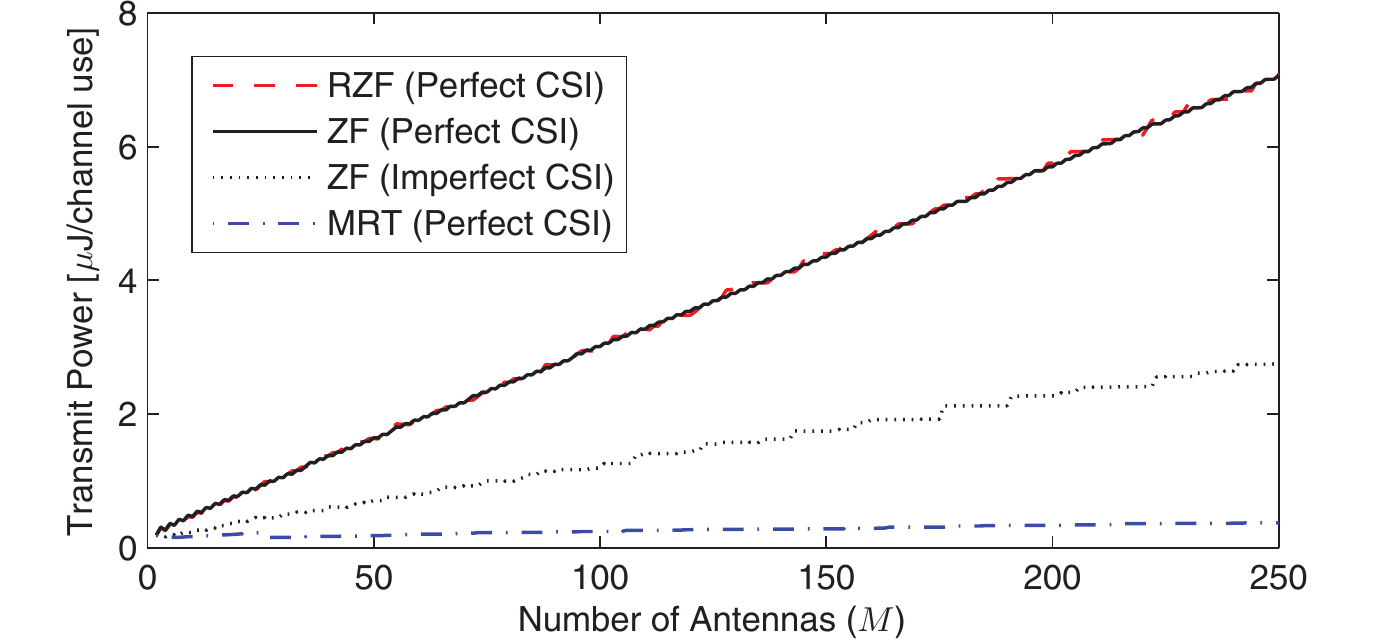}
\end{center} \vskip-4mm
\caption{EE-maximizing transmit power for different number of BS antennas. ZF (Perfect CSI) was computed analytically and the rest by MC simulations.} \label{figure-optpower} \vskip-1mm
\end{figure}

Fig.~\ref{figure-optpower} shows the transmit power that maximizes the EE for different $M$ (using the corresponding optimal $K$). We consider three precoding schemes with perfect CSI: ZF, regularized ZF (RZF), and MRT. We also show ZF with imperfect CSI acquired from uplink MMSE estimation  \cite{Bjornson2014a}. In all cases, the most energy efficient strategy is to \emph{increase} the transmit power with $M$. This is in line with Corollary \ref{cor:increase-power-with-M} but stands in contrast to the results in \cite{Hoydis2013a,Ngo2013a} which indicated that the transmit power should be decreased with $M$. The similarity between RZF and ZF shows an optimality of operating at high SNRs.

\begin{figure}[t]
\begin{center}
\includegraphics[width=\columnwidth]{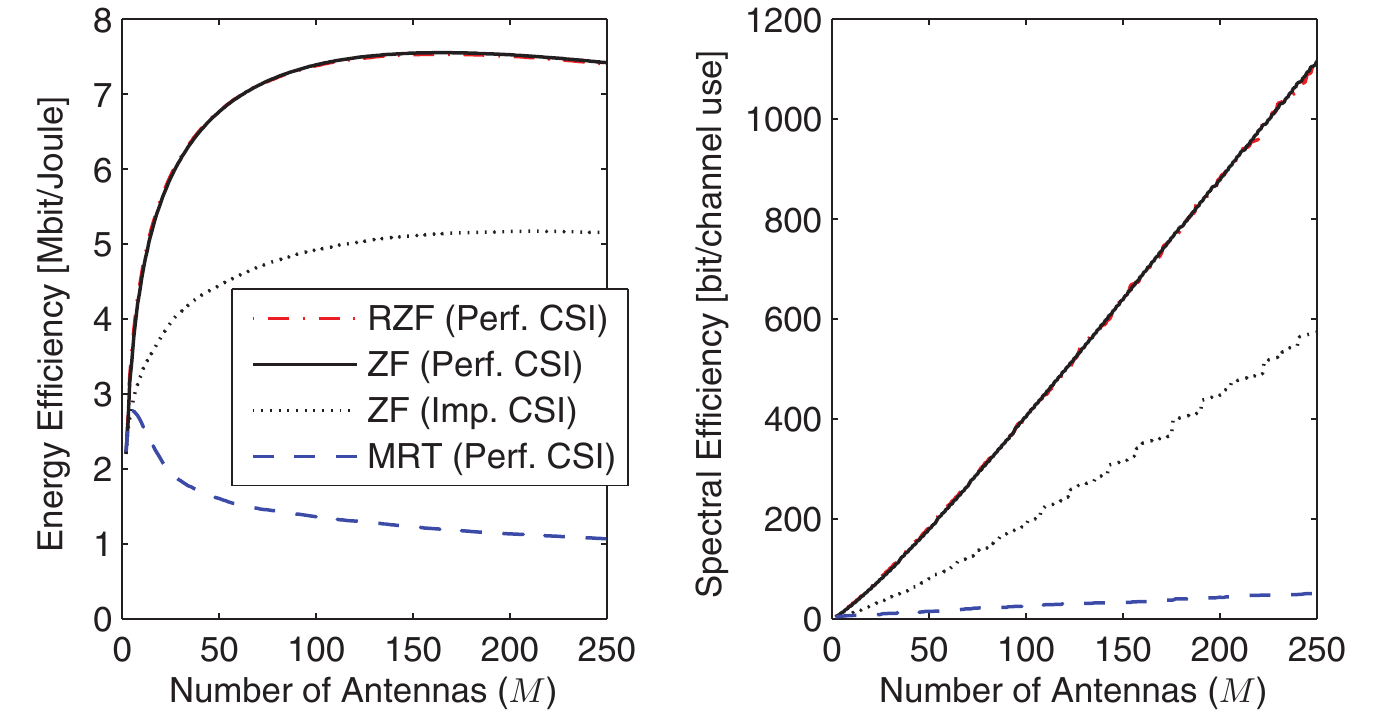}
\end{center} \vskip-4mm
\caption{Maximal energy efficiency and the corresponding spectral efficiency for different number of BS antennas and different precoding schemes.} \label{figure-optee-rates} \vskip-6mm
\end{figure}

Finally, Fig.~\ref{figure-optee-rates} shows the maximum EE for different number of BS antennas and the corresponding spectral efficiencies. We consider the same precoding schemes as in the previous figure. Firstly, we see that ZF performs similarly to the close-to-optimal RZF scheme. Secondly,  Fig.~\ref{figure-optee-rates} shows that there is a 3-fold difference in optimal EE between RZF/ZF and MRT under perfect CSI, while there is a 100-fold difference in spectral efficiency at the EE-maximizing operating points. The majority of this huge gain is achieved also under imperfect CSI, which shows that massive MIMO with proper interference-suppressing precoding can achieve both great energy efficiency and unprecedented spectral efficiencies.

\section{Conclusions}

This work analyzed how to select the number of BS antennas $M$, number of UEs $K$, and (normalized) transmit power $\rho$ to maximize the EE in the downlink of multi-user MIMO systems. Contrary to most prior works, we used a realistic power consumption model that explicitly describes how the total power depends on $M$, $K$, and $\rho$. Simple closed-form scaling laws for the EE-maximizing parameter values were derived under ZF precoding with perfect CSI and verified by simulations for other precoding schemes and imperfect CSI.

The EE (in bit/Joule) is a quasiconcave function of $M$ and $K$, which have therefore finite global optima.
Our numerical results show that deploying hundreds of antennas to serve a relatively large number of UEs is the EE-optimal macro-cell solution using today's circuit technology. Contrary to common belief, the transmit power should \emph{increase} with $M$ (to compensate for the increasing circuit power) and not decrease. Energy efficient systems are therefore \emph{not} operating in the low SNR regime and MRT should not be used. In other words, massive MIMO is an answer to the EE issues of current cellular networks, but only if proper interference-suppressing precoding (e.g., ZF or RZF) is applied.

The numerical results appear to be stable to small changes in the circuit power coefficients, but can otherwise change drastically. One can expect the circuit coefficients to decrease over time, which implies that the transmit power at the EE-optimal operating point also decreases. In the meantime, the most urgent bottlenecks might be to construct reliable large antenna arrays and find processors that are both highly energy efficient and fast enough to compute ZF/RZF almost instantly.

\appendices

\section*{Appendix: Proof of Lemma \ref{lemma:optimization-EE}} \label{app:lemma:optimization-EE}

The objective $g(z) = \frac{f \log_2(a+bz) }{c+d z}$ is quasiconcave since the level sets $S_{\alpha} =  \{ z \, : \, g(z) \geq \alpha, \, z > -\frac{a}{b} \} = \{ z \, : \, \alpha (c+d z) - f \log_2(a+bz) \leq 0 \}$ are strictly convex for any $\alpha \in \mathbb{R}$ \cite[Section 3.4]{Boyd2004a}. If there exists a point $z^{\mathrm{opt}}$ such that $g'(z^{\mathrm{opt}}) = 0$, the quasiconcavity implies that $z^{\mathrm{opt}}$ is the global maximizer and that the function is increasing for $z < z^{\mathrm{opt}}$ and decreasing for $z> z^{\mathrm{opt}}$.
To prove the existence of $z^{\mathrm{opt}}$, we note that \vskip-3mm
\begin{equation*}
g'(z) = 0 \quad \Leftrightarrow \quad \frac{1}{\log_e(2)}\frac{b  ( c+dz ) }{a+bz}  - d \log_2(a+bz) =0
\end{equation*} \vskip-1mm
\noindent by removing the non-zero denominator of $g'(z)$ and terms that cancel out.
This condition can be further rewritten as \vskip-2mm
\begin{equation} \label{eq:lemma-opt-problem-proof2}
\frac{ ( bc -ad ) }{a+bz} =  d \left( \log_e(a+bz) -1 \right).
\end{equation} \vskip-1mm
The substitution $x = \log_e(a+bz) -1$ transforms \eqref{eq:lemma-opt-problem-proof2} into $\frac{ ( bc -da ) }{de} \!=\! x e^x$.
The solution \eqref{eq:lemma-opt-problem-solution} is obtained since $z \!=\! \frac{e^{x+1} -a}{b}$.

\enlargethispage{2mm}

\bibliographystyle{IEEEtran}
\bibliography{IEEEabrv,refs}

\end{document}